# A spiking-domain implementation of electronic structure theory

Aakash Yadav[1†*], Daniel Hedman[2], Hongsik Jeong[1]

1) Graduate School of Semiconductor Materials and Devices Engineering, Ulsan National Institute of Science and Technology, 50 UNIST-gil, Eonyang-eup, Ulju-gun, Ulsan 44919, South Korea
2) Center for Multidimensional Carbon Materials (CMCM), Institute for Basic Science (IBS), Ulsan 44919, South Korea

**Abstract**

Electronic Structure Theory (EST) describes the behavior of electrons in matter and is used to predict material properties. Conventionally, this involves forming a Hamiltonian and solving the Schrödinger equation through discrete computation. Here, a new perspective to EST is provided by treating a perfectly crystalline material as a Linear Translation Invariant (LTI) system. The validity of this LTI-EST formalism is demonstrated by determining band structures for a one-dimensional chain of atoms, including the phenomenon of band structure folding in super cells. The proposed formalism allows for analytical traceability of band structure folding and offers computational advantage by bypassing the $O(N)$ eigenvalue calculations. The spike-based computing nature of the proposed LTI-EST formalism is highlighted; thereby implying potential for material simulations solely in the spiking domain.

**Keywords**: Electronic Structure Theory (EST), Linear Translationally Invariant (LTI), materials simulation, neuromorphic computing, Spiking Neural Networks (SNNs)

[†] A.Y. is now with Micron Technology Inc., Hyderabad, India
[*] Corresponding author email address: aakash.yadav@unist.ac.kr

**Introduction**

Electronic structure theory (EST) describes the behavior of electrons in materials and plays a vital role in understanding device physics. For instance, the electronic band structure of a material can provide insight into the relationship between the width of the first conduction band minimum and the effective mass of electrons. Specifically, a narrower first conduction band minimum corresponds to a lower effective mass of electrons, which in turn leads to higher electron mobility within the device [1]. To obtain such electronic band structures, several methods have been developed such as tight binding (TB) theory [1] which is a semi-empirical method, primarily used to calculate the single-particle Bloch states of materials. It relies on experimental data or first-principles calculations to parameterize the TB-Hamiltonian. First principles-based methods, on the other hand, solve the electronic structure of a material without the need for parametrization.

Recently, neural network-based approaches have been suggested for a plethora of problems in materials science [2], including the parametrization of the TB-Hamiltonian [3]. However, upon closer examination of these deep learning models, often referred to as "black boxes", trained on vast datasets, it becomes apparent that the underlying algorithms have their roots in the field of signals and systems. For example, it is the operation of convolutions that drives the state-of-the-art convolutional neural networks (CNNs). This is the same convolution integral that forms the basis for linear translation invariant (LTI) systems [4]. The use of CNNs to achieve superhuman performance in certain images tasks is also well reported [5]. Yet, the power consumption of these large neural networks is orders of magnitude higher than the human brain [6] which highlights the importance of spike-based computing and its connection with neuromorphic (or brain-inspired) computing.

To this end, Schuman et al. highlight the need for development of algorithms specific to neuromorphic hardware alongside the development of such hardware itself [7]. They particularly emphasize that neuromorphic computers, which promise computation with ultralow power consumption owing to their spike-domain computing, also require the adaptation of conventional algorithms to such spike-based computing. Additionally, Date et al. underline that algorithms for such computers need not be machine-learning specific [8]. Rather, their work proves the Turing-completeness of neuromorphic computing; thus, implying that all computation feasible on today's computers should be so on a neuromorphic computer too with the additional benefit of ultralow power consumption, thanks to its spike-based computing. Hence, materials simulation should be no exception. In fact, there is a broad spectrum of non-cognitive applications that neuromorphic computers aim to cater, as Aimone et al. highlight [9].

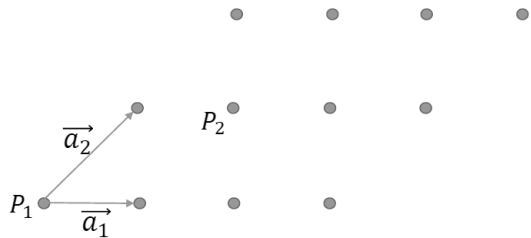

Figure 1: A 2D crystalline lattice. Here the arrows represent lattice vectors, $\vec{a_1}, \vec{a_2}$, and the dots represents lattice sites, $P_1, P_2$.

In this work, a new approach to solving EST problems is proposed based on LTI systems. Calculations based on this LTI-EST formalism are fully generic whereas the semi-empirical TB theory is employed for expansion of atomic orbitals, see Supplementary Material. Note that in choosing TB theory for such expansion, generality is not lost. Indeed, prior works have demonstrated the application of Green's function and Wannier orbitals to determine band structures through first principles [10]. Here, the equivalence of the LTI-EST formalism is illustrated by determining the band structure of a linear chain of atoms. The phenomenon of band structure folding is also explored, for which it has been demonstrated that concepts from signal analysis are important to the unfolding of band structures [11]. In the LTI-EST formalism a perfectly crystalline material is considered as an LTI system. It is known that for an LTI system, response to any unknown signal (perturbation) can be determined if the impulse response is known [4] & such determination becomes even simpler in the Fourier space because of the equivalence of convolution in real space to multiplication in Fourier space. A new way to derive band structures can be developed based on this insight.

**The LTI-EST Formalism**

The proposed LTI-EST formalism is derived from the Schrödinger equation with energy eigenvalue, $E$, and the corresponding eigenfunction, $\psi$, with $H$ being the Hamiltonian operator

$$H\psi = E\psi \rightarrow \left(-\frac{\hbar^2 \nabla^2}{2m^*} + U\right)\psi = E\psi. \qquad (1)$$

Here, $\hbar$ is the reduced Planck constant, $m^*$ the effective mass of electrons and $U$ the potential energy of the system. In the case of a 1D system of atoms, the Laplacian in the left-hand side of Equation 1 can be expanded as $-\frac{\hbar^2 \nabla^2 \psi}{2m^*} = -\frac{\hbar^2}{2m^*} \frac{\partial^2 \Psi}{\partial x^2}$.

To solve Equation 1 numerically, a Taylor series expansion can be used thus allowing for a finite difference approximation employing central difference method as follows, $\frac{\partial^2 \Psi}{\partial x^2}\big|_{x=x_n} = \frac{\psi(x_{n+1}) - 2\psi(x_n) + \psi(x_{n-1})}{a^2}$, which assuming $t_0 = \left(\frac{\hbar^2}{2m^* a^2}\right)$ and $(x_{n+1} - x_n) = (x_n - x_{n-1}) = a$, yields

$$\sum_m \left[-t_0 \delta_{n,m+1} + (2t_0 + U_n)\delta_{n,m} - t_0 \delta_{n,m-1}\right] \psi_m = E\psi_n \qquad (2)$$

It must be noted that the Hamiltonian in Equation 2 can be written as a circulant matrix. Thus, Equation 2 can be rewritten as a convolution owing to the equivalency between circulant matrices and the convolution operator [12]. Here, one must also note that due to the symmetrical nature of the kernel $[-t_0 \quad 2t_0 + U_n \quad -t_0]$, convolution & correlation would yield the same result since flipping it would result in no change; thus, implying equivalence between both the operations. This, in essence, implies that solving the Schrödinger equation is equivalent to a convolution operation once the right kernel is determined. Note that while calculations are only shown for the case of one dimension, this would also hold true for higher dimensions; thus, resulting in no loss of generality. As shown in Equation S1 of the Supplementary Material, for such a convolution operator, the Fourier Transform of the kernel gives the corresponding eigenvalues. The parameters $t_0$ and $U_n$ are material specific and can be determined either through experimental observations or ab initio calculations, similar to TB theory.

So far, all calculations have been generic and not specific to any particular method for the expansion of the atomic orbitals. In Section 2 of the Supplementary Material, such expansion is done using TB theory to arrive at Equation S2 without loss of generality. Now we show how band structures can be derived using the LTI-EST formalism in contrast to the conventional perspective presented in Section 2 of Supplementary Material. By leveraging concepts from signals analysis, in particular eigenvalue decomposition through Fourier transform, energy eigenstates are obtained resulting in the band structure. A perfect crystalline material can be considered as an LTI system having translational invariance instead of time invariance. As observed in Figure 1 all the lattice points of a perfect crystalline material can be traversed through the two vectors $\vec{a_1}$ and $\vec{a_2}$ and therefore, these two vectors span the entire lattice space. From the two points $P_1$ and $P_2$, although seemingly different, the lattice is the same implying translational invariance.

To prove linearity, one can realize that any lattice point $P$ in Figure 1 can be written as: $m\vec{a_1} + n\vec{a_2}$ where $m$ and $n$ can be any real numbers depending on the frame of reference. Let an operator $F$ operate upon this point $P$. It is assumed, at this point, that such an operator is of the form represented in Equation 3 below which is a general constant coefficient differential equation describing an LTI system [4]

$$a_n \frac{d^n y}{dt^n} + a_{n-1} \frac{d^{n-1} y}{dt^{n-1}} + \cdots + a_1 \frac{dy}{dt} + a_0 y(t) = b_n \frac{d^n x}{dt^n} + b_{n-1} \frac{d^{n-1} x}{dt^{n-1}} + \cdots + b_1 \frac{dx}{dt} + b_0 x(t). \quad (3)$$

Since $m$ and $n$ are constant real numbers and do not play any role in the differential operator such as in Equation 3, these can be factored out resulting in $F\left(m\vec{a_1} + n\vec{a_2}\right) = mF\left(\vec{a_1}\right) + nF\left(\vec{a_2}\right)$, proving linearity.

Now that the conditions of linearity as well as translational invariance have been proven for the perfect crystalline lattice, the LTI-EST formalism can be derived starting by rewriting Equation S2 as

$$E(k) = \sum_{R''} e^{\overrightarrow{jk(R'')}} \int \phi_s^*(x) H \phi_s(x - R'')(dx) = \sum_{R''} e^{-\overrightarrow{jk(R'')}} \int \phi_s^*(x) H \phi_s(x + R'')(dx). \quad (4)$$

Note the change of sign in Equation 4, here equivalence is achieved because the summation runs over all possible distances, $R''$, thus it does not matter if the sum is done from minus infinity to plus infinity or vice versa.

The right-hand side of Equation 4 can be interpreted as a Fourier transform of a cross-correlation operation. To be more specific, this cross-correlation is operated upon the atomic orbital itself and the resultant of Hamiltonian applied atomic orbital. Furthermore, cross-correlation is equivalent to convolution if the kernel used is symmetric and the result of both the operations (convolution & cross-correlation) is then equivalent. The derivations presented in this section indeed predict the symmetry of such a kernel.

**1D Chain of Atoms**

Viewing the crystal as an LTI system allows it to be characterized by its impulse response and since we establish an equivalence between band structures from TB theory and the LTI-EST formalism, the impulse response matrix must include all atomic neighbors required by TB theory. To find the impulse response, the intuitive first step is to begin with a unit impulse function. Depending on the material, this can be replaced with a 1D, 2D or even a 3D impulse function. Here, a 1D chain of atoms is considered thus a 1D unit impulse is sufficient. These impulse functions can be interpreted as unit cells where the $\delta$ functions are placed at the center of these unit cells. Clearly, such a placement can only hold true when the transport as well as transverse directions of the k-space are equivalent in two (or more) dimensions so that the unit cell can be replaced with an effective atom [1]. This necessarily implies that the first conduction band minimum (which primarily governs the electron mobility) should be same when observed along both these directions. In the case of 1D, this requirement will always be met. Using the analogy described herein, the band structures for all the super cells discussed in the Supplementary Material can be achieved more efficiently and with analytical insights. Henceforth, the unit cells (primitive or super cells) will be considered as input signals in space coordinates and the resulting band structures will be obtained as the response of the LTI system.

### A. Primitive cell of lattice constant $a$

For the case of 1D chain of atoms, a 1D $\delta$ function, placed at the center of the primitive cell (PC), describes the system. As shown in Figure 2 (b), the origin of the frame of reference is placed where the $\delta$ function lies (i.e., center of the PC). Therefore, mathematically, the primitive cell can be written as $\delta[x] = \begin{bmatrix} 0 & 1 & 0 \end{bmatrix}$.

To be consistent with the calculations presented in the Supplementary Material, only the nearest neighbor interactions are taken into consideration with $\alpha$ being the on-site energy and $\beta$ being the hopping parameter to account for such nearest neighbors in this 1D chain. The impulse response matrix $h[x]$ can then be written as

$$h[x] = \begin{bmatrix} \beta & \alpha & \beta \end{bmatrix}. \qquad (5)$$

Interestingly, the equivalence of this matrix with the Green's function should be highlighted at this point. Green's function, which is used for a variety of problems, holds its primary importance for solving inhomogeneous linear differential operator [13]. However, when the operator is LTI, the Green's function becomes equivalent to the impulse response. Thus, impulse response is a special case of Green's function. Upon taking Fourier transform of the Equation 5, $F(h[x]) = H[k_x] = \alpha + 2\beta \cos(ak_x)$, the result can be interpreted as the Green's function in the reciprocal space. One must also note that this is the same equation as Equation S3 in the Supplementary Material derived from TB theory. The resulting band structure plotted in Figure 2 (a) is in good agreement with the result obtained from the discretized calculation in Section 3.1 of the Supplementary Material.

### B. Super cells of lattice constants $2a$, $3a$ and $4a$

Results for super cells with varying lattice constants are presented next. Compared to primitive cells, such supercells are larger and therefore have a smaller Brillouin zone leading to folding of the band structure [1]. Symmetry is maintained throughout this work, meaning that Peierl's distortion is not considered [14]. For a super cell with $2a$ lattice constant, the new input signal is $i[x] = \begin{bmatrix} 0 & 1 & 1 & 0 \end{bmatrix}$.

As the concerned linear chain of atoms is an LTI system, its response is obtained through convolution with $h[x]$ as $o[x] = i[x] * h[x] = \begin{bmatrix} 0 & 1 & 1 & 0 \end{bmatrix} * \begin{bmatrix} \beta & \alpha & \beta \end{bmatrix} = \begin{bmatrix} \beta & \alpha+\beta & \alpha+\beta & \beta \end{bmatrix}$.

The same calculations can be done in analytical form instead of matrices. Figure 3 (d) represents the frame of reference. With this reference, the new input (physically implying a super cell of side $2a$) is: $i[x] = \delta[x] + \delta[x - a]$. Thus, the super cell becomes a linear summation of two $\delta$ functions. This frame also ensures that one unit cell still lies at the origin. Since these $\delta$ functions are placed at the center of the unit cells, it becomes clear that in a super cell, the center-to-center distance between any two such consecutive unit cells must be $a$.

The output of LTI system for this input is $o[x] = h[x] * i[x] = h[x] * (\delta[x] + \delta[x - a]) = h[x] + h[x - a]$.

To obtain energy eigenvalues, Fourier transform of a convolution integral (between Hamiltonian applied atomic orbital and the atomic orbital itself) needs to be performed. The Fourier transform of the convolution gives

$$E[k_x] = F(o[x]) = H[k_x](e^{j(0)} + e^{j(-k_x a)}). \quad (6)$$

The expression must be left as it is because Fourier transform diagonalizes the convolution operator whereby Fourier transform of the kernel itself yields the eigenvalue & the complex exponential yields the eigenfunction (see Equation S1 of the Supplementary Material). According to Bloch's theorem, only a certain set of wave vectors allow for the periodicity in Bravais lattice; thus, k-points in the reciprocal space only exist where $k_{x_{PC}} = \frac{2N\pi}{a}$ in PC in a simple cubic cell [1]. It should be noted that the discretization phase factor will vary accordingly with the geometry if the translation and transverse directions aren't the same. However, super cells (SCs) yield, $k_{x_{SC}} = \frac{2N\pi}{2a} = \frac{k_{x_{PC}}}{2}$ for integer values of $N$. Since these complex exponentials are also the eigenvectors, multiplication by a 'constant' eigenvalue (Fourier transform of the kernel) implies translation in k-space necessarily implying a phase shift. But by what amount? In Equation 6, the first eigenvector doesn't account for any phase shift; however, the second one shifts the phase by $ak_x$ but because of discretization $ak_x = 2\pi$ for any set of neighboring atoms. For a SC with a $2a$ lattice constant, this shift is divided by 2 (since $k_{x_{SC}} = \frac{k_{x_{PC}}}{2}$). Therefore, this results in two equations with which one can obtain the band structure. Plotted as solid lines in Figure 3 (a), this band structure agrees well with the calculations performed in Section 3.2 from the Supplementary Material.

$$\alpha + 2\beta \cos\left(\frac{i\pi}{2} + \frac{ak_x}{2}\right) \text{ with } i \in \{0,2\} \quad (7)$$

Similarly, for super cells with lattice constant $3a$ and $4a$, calculations can be carried out as follows:

$i_3[x] = \delta[x + a] + \delta[x] + \delta[x - a]$ and $i_4[x] = \delta[x + a] + \delta[x] + \delta[x - a] + \delta[x - 2a]$ with the used frame of references provided in Figure 3 (e) and Figure 3 (f) respectively. The outputs after convolution then yield: $o_3[x] = h[x + a] + h[x] + h[x - a]$ and $o_4[x] = h[x + a] + h[x] + h[x - a] + h[x - 2a]$ respectively for the two super cells. Using Fourier transform to find eigenvalues results in $E_3[k_x] = F(o_3[x]) = H[k_x](e^{j(k_x a)} + e^{j(0)} + e^{j(-k_x a)})$ & $E_4[k_x] = F(o_4[x]) = H[k_x](e^{j(k_x a)} + e^{j(0)} + e^{j(-k_x a)} + e^{j(-2k_x a)})$ for the two super cells respectively.

This results in three and four equations (given below) for the two SCs plotted with solid lines in Figure 3 (b) and Figure 3 (c) respectively which agree well with the band structures (plotted with red dots) obtained through discrete calculations using conventional perspective in Section 3.2 from Supplementary Material.

$$\alpha + 2\beta \cos\left(\frac{i\pi}{3} + \frac{ak_x}{3}\right) \text{ with } i \in \{-2,0,2\} \quad (8)$$

$$\alpha + 2\beta \cos\left(\frac{i\pi}{4} + \frac{ak_x}{4}\right) \text{ with } i \in \{-2,0,2,4\} \quad (9)$$

The reason why all the band structures in this work are presented in the domain of $[0, \pi]$ is to maintain uniformity with the published literature. One can, for example, refer to established literature [15, 16] and find consistency for the respective cases dealt with as part of this work.

Table I: Quantitative comparison of the new and conventional perspectives to band structure folding.

| Property | Conventional perspective | This work |
|---|---|---|
| Computational cost | $O(N)$ (Diagonalization) | $O(1)$ (Analytical) |
| Medium of modeling the cell | Atomic | Spikes/Impulse |
| Mode of computation for band structure folding | Discrete (Hermitian Hamiltonian) | Analog (Analytic equation) |
| Traceability of the folded band structure | No | Yes |
| SNN friendliness | No | Yes |

**Discussion**

A benefit of the LTI-EST formalism is the advantage it offers in terms of computational cost. As seen in Figure S1, an $O(N)$ loop is needed to obtain the band structure using the conventional perspective. This is a necessity because it requires one to iterate over all the k-points in the reciprocal space to find the set of eigenvalues and eigenvectors. In contrast to this, the new perspective "bypasses" the $O(N)$ loop as it is governed by the analytical insights obtained through eigenvalue decomposition using the complex exponential basis which diagonalizes the convolution operator. Superficially, the equation which is obtained for the case of a primitive cell is used for the cases of all the other super cells as well with the only difference being the phase shift and the change in the sampling rate in the analytical equation. Hence, the computational load effectively equates to $O(1)$ in the LTI-EST formalism.

This also makes one realize that the proposed perspective allows for analytically tracing the process of band structure folding. One can see, for example, in Figure 3 that the contribution from each of the neighboring cells can be traced down analytically in each band structure (color coded solid lines). Interestingly, all the band structures presented in Figure 3 which are obtained from the LTI-EST formalism solely rely on modeling the linear chain of atoms as spikes. It is noteworthy that in line with expectations from previous works [6-9], the proposed approach makes well-established EST compatible with spike-based computing; especially, addressing signal processing as a potential non-cognitive application of neuromorphic computing for materials simulation [9]. Since the proposed perspective completely operates in the spike-based computing domain, this implicitly implies a possibility of making materials simulation friendlier for Spiking Neural Networks (SNNs). Furthermore, since such SNNs inherently take temporal correlation into account & are event driven (by the timing of when a spike is sent), this implies the possibility of taking the time domain into consideration too through this perspective.

The LTI-EST formalism primarily draws its origin from the derivation done in Section 1 of Supplementary Material whereby the diagonalization of the convolution operation is achieved through Fourier transform. This "eigen" relation developed between the convolution operation and Fourier transform allows for the realization of complex exponential as the eigenvector- which makes it possible to then arrive at the eigenvalues analytically even for the super cells. The analysis made in this section can be summarized in the form of Table I. Besides the case of 1D chain of atoms discussed in this manuscript, the implementation of this LTI-EST formalism has also been presented to be effective on a more complex example of monolayer $MoS_2$ and the process flow is described in detail in this Master Thesis [17].

It must be mentioned, at this point, that although Green's function has been employed in prior literature to obtain band structures through ab initio methods too whereby such task was achieved through transformation from Wannier orbitals to k-space [9]; yet the novelty of this work lies in using an empirical method instead and introducing spike-domain computing. Besides, the LTI perspective wasn't exploited to the extent presented in this work.

It must also be noted that throughout this work, only a single orbital was employed. It might be interesting to further expand this framework to many orbitals. Moreover, another interesting application of this work can be in the understanding of amorphous materials or heterojunctions. However, for such studies, concepts of non-linear & translational variant systems might be required to be invoked as well.

**Conclusion**

In this work, a new perspective to the electronic structure theory is proposed employing concepts from the analysis of signals and systems. The effectiveness of this LTI-EST formalism is presented via demonstration of band structure folding for a 1D chain of atoms with varying lattice constants using conventional as well as the new perspectives. Moreover, the computational advantage and analytical insights offered by the proposed formalism have been highlighted. The significance of the spike-based computing offered by the proposed formalism is also reported to conclude with. Interestingly, this work succeeds in implementing a well-established theory in the spiking-domain which can be especially beneficial for futuristic neuromorphic computers.

**Acknowledgement**

This work was supported by the National Research Foundation of Korea (NRF) grant funded by the Korean government (MSIT) (No. 2021R1C1C2012077 and No. 2022M3I7A2079098).

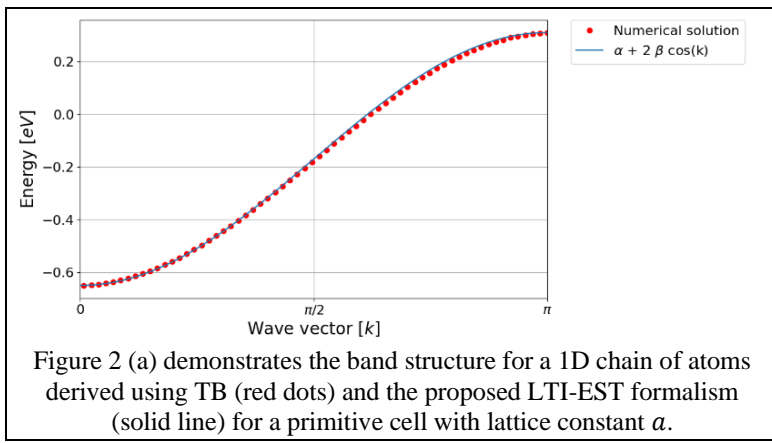
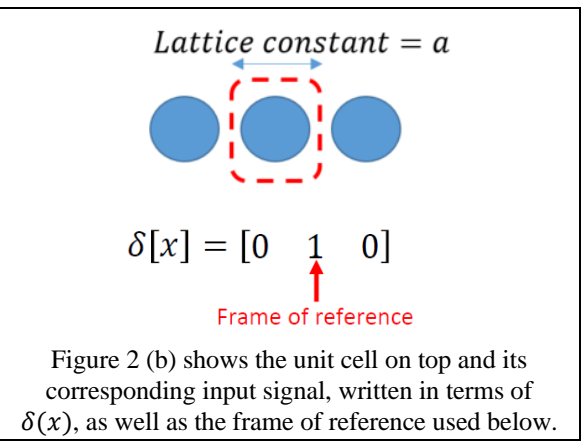

Figure 2 (a) demonstrates the band structure for a 1D chain of atoms derived using TB (red dots) and the proposed LTI-EST formalism (solid line) for a primitive cell with lattice constant $a$.

Figure 2 (b) shows the unit cell on top and its corresponding input signal, written in terms of $\delta(x)$, as well as the frame of reference used below.

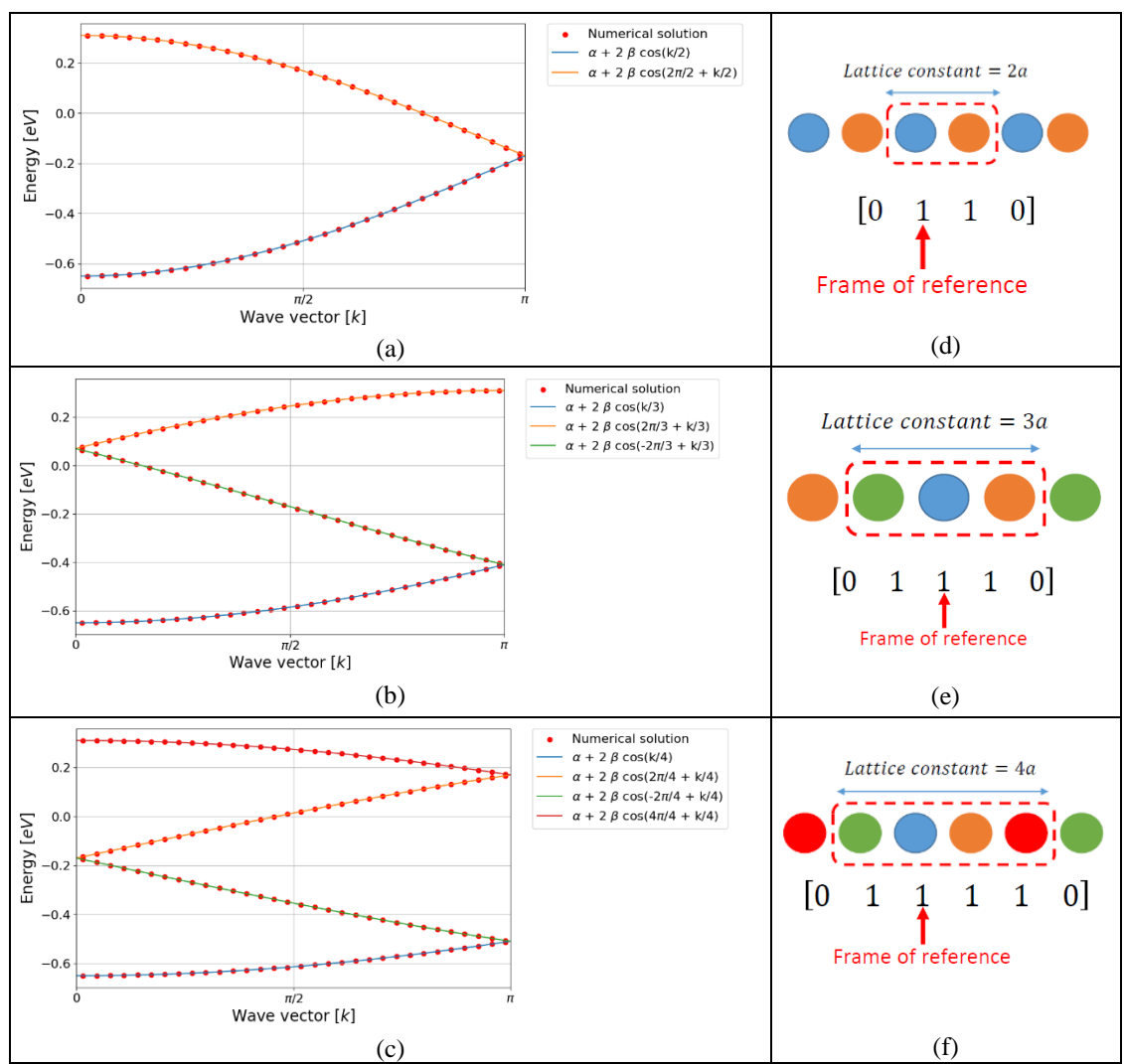

Figure 3: Super cells with lattice constant $2a$, $3a$ and $4a$ are depicted above in (d), (e) and (f) and the corresponding input signals, written in terms of $\delta(x)$, as well as the frame of reference used are shown below therein respectively. The band structures obtained using TB theory (red dots) and the proposed LTI-EST formalism (color-coded solid lines) are presented in (a), (b) and (c) respectively.

# Supplementary text to: "A spiking-domain implementation of electronic structure theory"

Aakash Yadav[1+*], Daniel Hedman[2], Hongsik Jeong[1]

1) Graduate School of Semiconductor Materials and Devices Engineering, Ulsan National Institute of Science and Technology, 50 UNIST-gil, Eonyang-eup, Ulju-gun, Ulsan 44919, South Korea
2) Center for Multidimensional Carbon Materials (CMCM), Institute for Basic Science (IBS), Ulsan 44919, South Korea

**Eigenvalue decomposition using Fourier Transform**

Eigenvalues and their corresponding eigenvectors can be calculated analytically for a matrix $A$ by equating the determinant of $(A - \lambda I)$ to 0. On a computer, diagonalization requires iterating over all data points of Hamiltonian matrix through $O(N)$ loop to find eigenvalues and eigenvectors (more details in Section 3 of the Supplementary Text). In the coming equations, it will be presented that the essential role of Fourier transform is to diagonalize the convolution operator. The transform indeed results in change of basis which also turns out to be the one in which convolution operator is diagonal; thereby, making it a lot easier to calculate the eigenvalues and eigenvectors.

The convolution operator represents the operation of linear superposition. Once the response of an LTI system to a delta function is known, the overall response may be obtained by taking the linear, translated & scaled superposition of all such responses which is what convolution operator does by smearing out a function via the 'response' of a linear system to a single $\delta$ function. The response to a $\delta$ function is termed as an impulse response [4]. Such an impulse response is a special case of Green's function (more details in Section 3.1 of the main text). But the convolution operator is not inherently diagonal, it mixes up coordinates. Nevertheless, in Fourier space, the convolution operator remains diagonal & doesn't mix up Fourier components. To mathematically prove this, convolution operator $G$ (with a kernel $h$) acting on a function $f(x)$ can be seen as $G[f(y)] = \int_{-\infty}^{+\infty} h(x)f(y-x)dx$. With $f(y) = e^{jky}$, equation then becomes:

$$G[f(y)] = G[e^{jky}] = \int_{-\infty}^{+\infty} h(x)e^{jk(y-x)}dx = e^{jky} \int_{-\infty}^{+\infty} h(x)e^{-jkx}dx = \lambda f(y) \quad \text{(S1)}$$

Here, an 'eigen' relationship has been demonstrated whereby the complex exponential has been proven to be the eigenvector and the Fourier transform of the kernel has been proven to be the eigenvalue of an LTI system (or the convolution operator in general). The brief derivation above shows that taking Fourier transform amounts to choosing a basis of the complex exponentials; and in this Fourier space, the convolution operator does not mix up Fourier components; it acts upon each component independently. An implementation of the theoretical background developed in this section is presented in the main text.

## The conventional perspective

In this section, TB theory would be employed and the single particle Schrödinger equation for the states will be solved for in a crystal by expanding the Bloch states in terms of LCAO [1]. For a crystal lattice, $\psi$ is referred to as the "single particle state". In simpler terms, it represents the energy states that are available within the crystal for the electrons to occupy. Please note that $H = H_{at} + \Delta U$. Here, $\Delta U$ encodes all the differences between the true potential in the crystal and the potential of an isolated atom. Since $H_{at}$ represents the atomic Hamiltonian, $H_{at} \phi = \varepsilon \phi$ where $\phi$ represents the atomic orbital wavefunction for isolated atom and is orthonormal, i.e.,

$$\int \phi_i^*(r)\phi_j(r+R)(dr) = \delta_{ij} = \begin{cases} 1 \text{ if } i = j \text{ and } R = 0 \\ 0 \text{ otherwise} \end{cases}$$

Hopping parameter determining the extent of overlap $= \gamma(|R|) = \int \phi_i^*(r) H \phi_i(r+R)(dr)$. Owing to the periodicity of the lattice potential, Bloch's theorem can be utilized at this point for the single particle state: $\psi_{n,k}(r+R) = e^{(\vec{jk})(\vec{R})} \psi_{n,k}(r)$. Although a single atomic orbital wave function would not satisfy this equation, owing to LCAO, a linear combination of them would:

$$\psi_{n,k}(r) = \frac{1}{\sqrt{N}} \sum_R \phi_n(r-R) e^{\vec{jk}\vec{R}}$$

$H$ is an operator and hence expressible as a matrix each of whose elements can be expressed as: $\langle i|H|j \rangle$ and since $H\psi = E\psi$, $E = \langle i|H|j \rangle$. Therefore, $E(k) = \int \psi_k^*(r) H \psi_k(r)(dr)$. In this equation, the expression for $\psi$ can be substituted again to express it in the form of individual atomic orbitals:

$$E(k) = \frac{1}{N} \sum_R \sum_{R'} e^{\vec{jk}(\vec{R'}-\vec{R})} \int \phi_s^*(r-R) H \phi_s(r-R')(dr)$$

$$E(k) = \frac{1}{N} \sum_R \sum_{R'} e^{\vec{jk}(\vec{R''})} \int \phi_s^*(x) H \phi_s(x-R'')(dx)$$

$$E(k) = \sum_{R''} e^{\vec{jk}(\vec{R''})} \int \phi_s^*(x) H \phi_s(x-R'')(dx) \tag{S2}$$

TB theory is based on the notion that the extent of interaction between two neighboring atoms decays with distance. Thus, as one of the very first steps, it is required to decide how many neighbors to account for beyond which the interaction potential would subside (i.e., considered zero for all calculations). For $R'' = 0$, $E(k) = \int \phi_s^*(x) H \phi_s(x)(dx) = \varepsilon_s$ (onsite energy). For $R'' \gg 0, E(k) = 0$. For nearest neighbour ($R'' = \tau$):

$$E(k) = \varepsilon_s + \sum_\tau e^{\vec{jk}(\vec{\tau})} \int \phi_s^*(x) H \phi_s(x-\tau)(dx) = \varepsilon_s + \sum_\tau e^{\vec{jk}(\vec{\tau})} \gamma(|\tau|)$$

# 1D chain of atoms

In this section, conventional perspective is implemented to obtain band structures of a 1D chain of atoms.

## A. Primitive cell of lattice constant $a$

In this case, one can simply proceed from Equation S2. Assuming the on-site energy to be $\alpha$ and the hopping parameter $\beta$ to account for only the nearest neighbors in this 1D chain, one can obtain the equation for energy eigenvalues. To begin with, a primitive cell of lattice constant '$a$' is considered. Pictorially, this is represented in the Figure 2 (b). Equation S2 can be written then as:

$$E(k) = \varepsilon_s + \sum_\tau e^{\overrightarrow{jk(\tau)}} \gamma(|\tau|) = \alpha + \beta(e^{jk_x a} + e^{-jk_x a}) = \alpha + 2\beta \cos(k_x a) \qquad (S3)$$

The above operation of eigenvalue calculation can also be done through diagonalization of the discrete Hamiltonian to obtain the band structure for this primitive cell through discrete operations as shown in Figure 2 (a) as red circles. For consistency, parameters $\alpha = -0.17\ eV$ & $\beta = -0.24\ eV$ are kept same throughout this work.

## B. Super cell of lattice constant $2a$

A super-cell of lattice constant $2a$ results in a bigger cell as compared to the primitive cell, i.e., a smaller Brillouin zone and thus, lead to folding of the band structure [1] (shown in Figure 3). As part of the conventional method of obtaining energy eigenvalues, individual Hamiltonian interaction matrices need to be formed before obtaining the final Hermitian Hamiltonian matrix the eigenvalues of which will yield the band structure. The individual interaction matrices for the super-cell of lattice constant $2a$ can be obtained as: $H_{-1,0} = \begin{bmatrix} 0 & \beta \\ 0 & 0 \end{bmatrix}, H_{0,0} = \begin{bmatrix} \alpha & \beta \\ \beta & \alpha \end{bmatrix}, H_{1,0} = \begin{bmatrix} 0 & 0 \\ \beta & 0 \end{bmatrix}$. Accounting for the exponential factors then, the Hermitian Hamiltonian matrix can be obtained as follows:

$$H = \begin{bmatrix} \alpha & \beta + \beta e^{jak_x} \\ \beta + \beta e^{-jak_x} & \alpha \end{bmatrix}$$

From this point, the eigenvalues and the eigenvectors need to be found for each valid value of $k_x$. In Figure S1, this process is shown for Python language. It can be easily observed that the process shown involves iterating over the set of all available k-points through an $O(N)$ loop. The process depicts finding eigenvalues and eigenvectors through diagonalization which is a discrete operation. Such values are calculated at each point and since the size of the matrix is 2x2; two eigenvalues are obtained at every point. The resulting band structure from this discrete operation is presented in Figure 3 (a) as red circles.

```python
import cmath, math
import numpy as np
from numpy import linalg

alpha, beta, ans1, ans2 = -0.17, -0.24, [], []

k = np.linspace(-4*math.pi, 4*math.pi, 256)

for i in range(0, 256):
    tmp = linalg.eig([
        [alpha, beta + beta*cmath.exp(complex(0, k[i]))],
        [beta + beta*cmath.exp(complex(0, -k[i])), alpha]
    ])[0]
    ans1.append(tmp[0].real)
    ans2.append(tmp[1].real)
```

Figure S1: Demonstration that calculation of eigenvalues requires $O(N)$ loop for the diagonalization at each point.

### C. Super cell of lattice constant $3a$

The Hermitian Hamiltonian matrix for the super-cell of lattice constant $3a$ can be obtained as follows. The case has been pictorially presented in the Figure 3 (e) with the resulting band structure demonstrated in the Figure 3 (b) as red circles.

$$H = \begin{bmatrix} \alpha & \beta & \beta e^{jak_x} \\ \beta & \alpha & \beta \\ \beta e^{-jak_x} & \beta & \alpha \end{bmatrix}$$

### D. Super cell of lattice constant $4a$

The Hermitian Hamiltonian matrix for the super-cell of lattice constant $4a$ can be obtained as follows. The case has been pictorially presented in the Figure 3 (f) with the resulting band structure demonstrated in the Figure 3 (c) as red circles.

$$H = \begin{bmatrix} \alpha & \beta & 0 & \beta e^{jak_x} \\ \beta & \alpha & \beta & 0 \\ 0 & \beta & \alpha & \beta \\ \beta e^{-jak_x} & 0 & \beta & \alpha \end{bmatrix}$$